# *CHEVRON*'S SLIDING SCALE IN *Wyeth v. Levine*, 129 S. Ct. 1187 (2009)

For years now, courts and commentators have struggled to reconcile the presumption against preemption—the interpretive canon that presumes against federal incursion into areas of traditional state sovereignty—with the *Chevron* doctrine, which instructs courts to defer to reasonable agency interpretations of ambiguous federal statutes. Last Term, in *Wyeth v. Levine*, the Court held that the Food and Drug Administration's (FDA) drug labeling requirements did not preempt a state law failure-to-warn claim against a drug manufacturer.[1] In so holding, the Court found no entitlement to deference for an FDA regulatory preamble in favor of preemption. This decision provides further fodder for those critics focused on the Court's long history of seemingly arbitrary reliance on agency input and haphazard application of the presumption against preemption.[2] A close examination, however, reveals both why the Court has been reluctant to apply an across-the-board standard of deference to agency views and what an appropriate framework for agency deference might look like. The Court's inconsistent approach to preemption cases results from its intense focus on congressional intent. A different approach would satisfy critics of the Court's inconsistency while allowing the Court to retain its focus on congressional intent: accord greater deference to agency views when Congress speaks clearly through an express preemption provision and lesser deference when Congress is silent.

---

1. 129 S. Ct. 1187 (2009).
2. *See, e.g.*, Mary J. Davis, *The Battle Over Implied Preemption: Products Liability and the FDA*, 48 B.C. L. REV. 1089, 1093–94 (2007) ("The proper weight of an agency's determination of preemption scope has generated much debate within the Supreme Court and among commentators. The Court has not answered the question of how an agency position affects the operation of implied conflict preemption doctrine, nor has it addressed how the historic primacy of state regulation in the area of health and safety is to be considered in the balance."); Paul E. McGreal, *Some* Rice *with Your* Chevron*?: Presumption and Deference in Regulatory Preemption*, 45 CASE W. RES. L. REV. 823, 826 (1995) ("While the Court has spoken on regulatory preemption, it has neither explained nor justified its position. Instead, the Court merely has applied *statutory* preemption rules to *regulatory* preemption cases. To the extent that statutory and regulatory preemption are different—under the Court's larger jurisprudence—difficulty may be expected in applying the same set of preemption rules to both areas."); Nina A. Mendelson, Chevron *and Preemption*, 102 MICH. L. REV. 737, 739 (2004) ("When faced with an agency interpretation addressing a statute's preemptive effect, courts have trod unevenly in reconciling *Chevron* deference with the *Rice* presumption against preemption."); Caleb Nelson, *Preemption*, 86 VA. L. REV. 225, 232–33 (2000) ("Most commentators who write about preemption agree on at least one thing: Modern preemption jurisprudence is a muddle."); Catherine M. Sharkey, *Products Liability Preemption: An Institutional Approach*, 76 GEO. WASH. L. REV. 449, 454 (2008) ("It is exceedingly difficult to demonstrate that any consistent principle or explanatory variable emerges from the Supreme Court's products liability preemption jurisprudence.").

In *Wyeth*, the Court considered the common law negligence claim of Diane Levine against the drug manufacturer Wyeth. Levine, suffering from a severe migraine headache, consented to a physician assistant's administration of Phenergan, a drug manufactured by the defendant.[3] The drug can be administered either intramuscularly or intravenously, and intravenous administration can be performed by either the IV-drip method or the faster but riskier IV-push method.[4] Because her symptoms were severe and an initial administration of the drug had failed to provide relief, the physician assistant administered the drug via the IV-push method, which promises faster relief but also carries a risk of significant side effects.[5] The drug is corrosive, and if it escapes from the vein into surrounding tissue it causes irreversible gangrene.[6] Unfortunately, in Levine's case this precise danger was realized. As the physician assistant administered the drug, it escaped the vein and came in contact with arterial blood, resulting in gangrene and eventually requiring the amputation of Levine's right forearm.[7] As a result of this amputation, Levine incurred substantial medical expenses and was forced to abandon her career as a professional musician.[8]

Levine brought a common law failure-to-warn claim against Wyeth alleging that Phenergan was defectively labeled.[9] Wyeth responded by arguing that federal law preempted Levine's negligence claim. Wyeth urged a finding of both impossibility and obstacle preemption. First, it argued that it would have been impossible for it to comply with a state common law duty to modify Phenergan's label while also remaining in compliance with FDA regulations. Second, Wyeth argued that recognition of the plaintiff's state tort action would create an unacceptable obstacle to the accomplishment of the purposes and objectives of Congress by substituting a lay jury's decision about drug labeling for the expert judgment of the FDA.[10] After a Vermont state trial court rejected Wyeth's motion for judgment as a matter of law on the preemption issue, a jury found Wyeth liable for negligence.[11] Although the drug's label warned of the danger of gangrene following inadvertent intra-arterial injection, its labeling was nonetheless defective because it failed to instruct clinicians to use the IV-drip method as an alternative to the riskier IV-push method.[12] The Vermont Supreme Court affirmed, holding that compliance with both federal and state law would have been possible and that common law liability posed no obstacle to the accomplishment of congressional objectives.[13]

---

3. *Wyeth*, 129 S. Ct. at 1191.
4. *Id.*
5. *Id.*
6. *Id.*
7. *Id.*
8. *Id.*
9. *Id.* at 1191–92.

The Supreme Court affirmed.[14] Justice Stevens, writing for the majority[15] and rejecting both theories of preemption, relied on two guiding principles: first, that "the purpose of Congress is the ultimate touchstone in every pre-emption case";[16] and second, that "the historic police powers of the States [are] not to be superseded by [a] Federal Act unless that was the clear and manifest purpose of Congress."[17] In response to Wyeth's impossibility preemption defense, the Court noted that although a manufacturer generally may not change its label after it is approved by the FDA, federal regulations do provide a mechanism[18] for manufacturers to change a label to add to or

---

10. *Id.* at 1193–94.
11. *Id.* at 1193.
12. *Id.* at 1191–92.
13. Levine v. Wyeth, 944 A.2d 179, 189–90 (Vt. 2006).
14. *Wyeth*, 129 S. Ct. at 1204.
15. Justice Stevens was joined by Justices Kennedy, Souter, Ginsburg, and Breyer.
16. *Wyeth*, 129 S. Ct. at 1194–95 (quoting Medtronic, Inc. v. Lohr, 518 U.S. 470, 485 (1996)).
17. *Id.* (quoting Rice v. Santa Fe Elevator Corp., 331 U.S. 218, 230 (1947)).
18. FDA regulations permit drug manufacturers to modify labels "[t]o add or strengthen a contraindication, warning, precaution, or adverse reaction [or to] add strengthen its warnings.[19] Because federal law permitted Wyeth unilaterally to strengthen the warnings on Phenergan's label, it was not impossible for Wyeth to comply with both federal and state requirements.[20]

The Court also rejected Wyeth's obstacle preemption argument. Wyeth argued that Congress intended FDA regulations to establish both a floor and a ceiling for drug labeling requirements and that to allow a state negligence cause of action would interfere with that objective.[21] The Court, however, found no evidence that Congress had expressed such an intent. Looking to the legislative history of the Food, Drug, and Cosmetic Act (FDCA), the Court noted that although Congress intended for the FDCA to bolster consumer protection against harmful products, Congress had intentionally provided no federal cause of action.[22] Instead, Congress had decided to rely on state tort law to provide appropriate relief. Such a policy, the Court reasoned, is inconsistent with the notion that Congress intended to preempt state common law.

In so deciding, the Court also rejected Wyeth's argument for deference to the FDA. In a preamble to a 2006 FDA regulation, the FDA declared that the FDCA should be read to establish both a floor and a ceiling for drug labeling, preempting conflicting state labeling laws.[23] The Court acknowledged that in earlier cases it accorded "some weight" to agency views regarding the impact of state law on federal objectives,[24] but maintained that it never deferred to an agency's *conclusion* that state law was preempted.[25] Rather, the weight accorded to

---

or strengthen an instruction about dosage and administration that is intended to increase the safe use of the drug product." 21 C.F.R. § 314.70(c)(6)(iii)(A), (C) (2009).

19. *Wyeth*, 129 S. Ct. at 1196.

20. *Id.* at 1199.

21. *Id.*

22. *See id.* at 1199–200 & n.7.

23. *Id.* at 1200; Requirements on Content and Format of Labeling for Human Prescription Drug and Biological Products, 71 Fed. Reg. 3922, 3934–35 (Jan. 24, 2006).

24. *Wyeth*, 129 S. Ct. at 1201; *see also, e.g.*, Geier v. Am. Honda Motor Co., 529 U.S. 861, 883 (2000) ("We place some weight upon [the agency's] interpretation of [the regulation's] objectives and its conclusion, as set forth in the Government's brief, that a tort suit such as this one would 'stan[d] as an obstacle to the accomplishment and execution' of those objectives.").

25. *Wyeth*, 129 S. Ct. at 1201.

agency views depends on their "thoroughness, consistency, and persuasiveness."[26] Applying this standard, the majority determined that the FDA's preamble should be accorded no deference. The FDA's notice of proposed rulemaking gave no indication of the regulation's potential federalism implications and did not offer interested parties an opportunity to comment, and the regulation itself reversed the agency's own longstanding position without providing an explanation for its decision.[27]

Justice Breyer concurred, emphasizing that the Court's decision did not reach the question of whether state tort law might interfere with and therefore be preempted by a future, specific regulation bearing the force of law.[28]

Justice Thomas, concurring only in the judgment,[29] agreed with the majority that FDA labeling requirements did not preempt Levine's common law negligence claim, but his rationale was markedly different. Rather than analyzing Wyeth's obstacle preemption argument under the Court's longstanding framework, Justice Thomas rejected as unsound the Court's entire line of obstacle-preemption jurisprudence.[30] Drawing on the theory of dual sovereignty in *Federalist* No. 51[31] and the Constitution's Bicameralism and Presentment Clause,[32] Justice Thomas emphasized his "increasing[] reluctan[ce] to expand federal statutes beyond their terms through doctrines of implied pre-emption."[33] Preemption, he argued, must turn on something more than "generalized notions of congressional purposes that are not contained within the text of federal law."[34] Preemption must instead be grounded in a direct conflict between state law and the text of validly promulgated federal law.[35]

---

26. *Id.* (citing Skidmore v. Swift & Co., 323 U.S. 134, 140 (1944)).

27. *Id.* at 1201–02.

28. If, for example, the FDA found state tort law to interfere with its desire to mandate a precise set of instructions on a drug label, the FDA's determination would potentially have preemptive effect. *Id.* at 1204 (Breyer, J., concurring).

29. *Id.* at 1204 (Thomas, J., concurring in the judgment).

30. *Id.* at 1211 ("This Court's entire body of 'purposes and objectives' preemption is inherently flawed. The cases improperly rely on legislative history, broad atextual notions of congressional purpose, and even congressional inaction in order to pre-empt state law.").

31. THE FEDERALIST NO. 51, at 323 (James Madison) (Clinton Rossiter ed., 1961) ("In the compound republic of America, the power surrendered by the people is first divided between two distinct governments, and then the portion allotted to each subdivided among distinct and separate departments. Hence a double security arises to the rights of the people. The different governments will control each other, at the same time that each will be controlled by itself.").

32. U.S. CONST. art. I, § 7, cl. 2 ("Every Bill which shall have passed the House of Representatives and the Senate, shall, before it become a Law, be presented to the President of the United States . . . .").

Justice Alito dissented.[36] He criticized the majority for taking an overly narrow view of obstacle preemption. "Congress made its 'purpose' plain in authorizing the FDA—not state tort juries—to determine when and under what circumstances a drug is 'safe,'"[37] he asserted, and by finding the plaintiff's state tort law claims unpreempted, the majority "upset the regulatory balance struck by the federal agency."[38] The dissent criticized the majority's treatment of the FDA's preemption determination, noting that in *Geier v. American Honda Motor Co.*[39] the Court had relied on materials other than a formal agency regulation and therefore could not be understood to have established a force-of-law test that would grant deference to a formal regulation but not to a preamble.[40] Finally, the dissent criticized the majority for its reliance on the presumption against preemption in the context of its obstacle preemption analysis, arguing instead that, as in *Geier*, the Court's analysis should have been driven by "'ordinary,' 'longstanding,' and 'experience-proved principles of conflict pre-emption.'"[41]

*Wyeth* stands as one more case in the Court's notoriously convoluted line of preemption decisions.[42] Particularly frustrating to commentators[43] is the Court's repeated failure to resolve

---

33. *See Wyeth*, 129 S. Ct. at 1207 (Thomas, J., concurring in the judgment) (quoting Bates v. Dow Agrosciences LLC, 544 U.S. 431, 459 (Thomas, J., concurring in the judgment in part and dissenting in part)).

34. *Id.*

35. *Id.* at 1208.

36. Justice Alito was joined by Chief Justice Roberts and Justice Scalia.

37. *Wyeth*, 129 S. Ct. at 1219 (Alito, J., dissenting).

38. *Id.* at 1220.

39. 529 U.S. 861 (2000).

40. *Wyeth*, 129 S. Ct. at 1228 (Alito, J., dissenting).

41. *Id.*

42. *See supra* note 2.

43. Commentators have proposed a number of frameworks to reconcile the tension between *Chevron* and the Court's preemption jurisprudence. *See, e.g.*, Mendelson, *supra* note 2 (suggesting that concerns of agency inexpertness, arbitrary decision making, and self-aggrandizement militate in favor of applying the lower explicitly and unambiguously the apparent tension between *Chevron* doctrine—requiring deference to an agency's reasonable interpretation of an ambiguous statute[44]—and the presumption against preemption, which requires a "clear statement" from Congress before a court may conclude that a federal statute preempts state law.[45] Does the federalism-inspired interpretive canon presuming against preemption serve as a "traditional tool of statutory construction"[46] resolving ambiguity and obviating the need for *Chevron* deference to agency views, or does the peculiar competence of agencies within their statutory spheres require deference even regarding such sensitive questions as preemption? And if agency views require deference, should that deference be tempered in light of countervailing federalism concerns?

At first glance, *Wyeth* adds little to the ongoing debate. The majority followed its recent precedent in refusing, without elaboration, to grant the agency's interpretation full-fledged *Chevron* deference, leaving the tension between the Court's preemption and *Chevron* decisions unresolved. Precisely what level of respect is due to agency interpretations remains unclear,[47] and the Court's view on the presumption against pre-

---

deference standard of *Skidmore v. Swift & Co.*, 323 U.S. 134 (1944), in the context of agency preemption determinations); Sharkey, *supra* note 2 (suggesting that although courts should not defer to agency *conclusions* regarding preemption, they should recognize agencies' superior ability to supply and analyze empirical data relevant to the desirability of a uniform federal regulatory system).

44. *See* Mendelson, *supra* note 2, at 743 ("Under the familiar rule of *Chevron*, unless Congress has directly answered the interpretive question at hand . . . a court must defer to an agency's reasonable interpretation of a statute it administers."); *see also* Chevron U.S.A., Inc. v. Natural Res. Def. Council, 467 U.S. 837, 844 (1984).

45. *See* Mendelson, *supra* note 2, at 752; *see also* Rice v. Santa Fe Elevator Corp., 331 U.S. 218, 230 (1947) ("[W]e start with the assumption that the historic police powers of the States were not to be superseded by the Federal Act unless that was the clear and manifest purpose of Congress.").

46. The *Chevron* framework requires courts to apply traditional tools of statutory construction to determine whether the text evinces clear congressional intent regarding the question at issue. Only if these tools fail to produce a clear meaning are courts to defer to agency interpretations. *See Chevron*, 467 U.S. at 843 n.9 ("If a court, employing traditional tools of statutory construction, ascertains that Congress had an intention on the precise question at issue, that intention is the law and must be given effect.").

47. *Compare, e.g.*, Geier v. Am. Honda Motor Co., 529 U.S. 861, 863 (2000) (according "some weight" to an agency's conclusion that state law would stand as an obstacle to federal goals), *and* Medtronic v. Lohr, 518 U.S. 470, 495 (1996) (noting emption is severely fractured.[48] By situating *Wyeth* within the larger framework of the Court's recent preemption decisions, however, it is possible to extract several fundamental principles that animate the Court's jurisprudence. These principles explain both why the Court has been reluctant to apply across-the-board *Chevron* deference to agency preemption determinations, and, perhaps more interestingly, what an appropriate framework for *Chevron*-style deference might look like in the preemption context.

First, the Court's decisions strongly emphasize the importance of congressional intent. In *Wyeth*, for example, the Court began its analysis with the "cornerstone" principle that "the purpose of Congress is the ultimate touchstone in every preemption case."[49] Later, considering the preemptive effect of the FDA's preamble, the Court refused deference to the agency's view because it conflicted with the Court's interpretation of congressional intent.[50] This focus on congressional intent results from an intense consciousness of the Court's role as a protector of federalism and perhaps also skepticism regarding agencies' competence to make significant national policy decisions outside their traditional realms of expertise.[51]

---

that the Court is "substantially informed by" any agency's interpretation of its regulations' preemptive scope), *with* Wyeth v. Levine, 129 S. Ct. 1187, 1201 (2009) ("[T]he weight we accord the agency's explanation of state law's impact on the federal scheme depends on its thoroughness, consistency, and persuasiveness."), *and* Riegel v. Medtronic, Inc., 552 U.S. 312, 329–30 (2008) ("Neither accepting nor rejecting the proposition that this regulation can properly be consulted to determine the statute's meaning").

48. The *Wyeth* majority relied on the presumption as a "cornerstone" of its decision, whereas the dissent claimed that the presumption is irrelevant in the context of conflict preemption. *Wyeth*, 129 S. Ct. at 1194–95 & n.3, 1228–29. Concurring in the judgment, Justice Thomas reserved the question of the applicability of the presumption, finding its resolution unnecessary given the clarity of the relevant statutes and regulations. *Id.* at 1208 n.2 (Thomas, J., concurring in the judgment).

49. *Id.* at 1194–95 (majority opinion).

50. *Id.* at 1201.

51. *See* Mendelson, *supra* note 2, at 755–56 (attributing the Court's use of the presumption against preemption to a "reluctance to risk incidental statutory interference with federalism values and with state sovereignty" and an "attach[ment] of substantive value to federalism goals"); *id.* at 779–91 (suggesting that agency determinations regarding preemption should not be accorded *Chevron* deference because agencies lack institutional competence to make such decisions).

Second, although the Court's selective application of the presumption against preemption initially appears haphazard, it, too, can be explained by the Court's focus on congressional intent. Critics have frequently noted that although the presumption factors strongly in many decisions,[52] it goes completely unmentioned in other seemingly analogous cases.[53] Indeed, in *Wyeth* itself the majority and the dissent were unable to agree on the presumption's proper role. The majority adopted the presumption as a "cornerstone" of its analysis, whereas the dissent deemed the presumption inapplicable to cases of implied preemption.[54]

Despite the Court's ambiguity, it is possible to say a few things about the presumption's role in preemption cases. Although the presumption's influence has gradually waned over the last few decades, it has retained its force in cases of implied preemption, where congressional intent is least certain.[55] Thus,

---

52. *See* Sharkey, *supra* note 2, at 506 ("[W]here [the presumption] rears its head, its effect is seemingly outcome determinative.").

53. *See id.* at 458 ("Here I join a veritable chorus of scholars pointing out the Court's haphazard application of the presumption. In the realm of products liability preemption, the presumption does yeoman's work in some cases while going AWOL altogether in others."); *see also* Calvin Massey, *"Joltin' Joe Has Left and Gone Away": The Vanishing Presumption Against Preemption*, 66 ALB. L. REV. 759, 759–64 (2003). *Compare, e.g.*, Medtronic, Inc. v. Lohr, 518 U.S. 470, 485 (1996) (applying the presumption to interpret the Medical Device Act), *with* Riegel v. Medtronic, Inc., 552 U.S. 312 (2008) (failing even to mention the presumption when interpreting the same statute).

54. *See supra* note 48. Although Justice Thomas in his concurrence did not decide the question of the presumption's applicability in the context of obstacle preemption, his complete rejection of the Court's prior obstacle preemption jurisprudence can be interpreted as an implicit adoption of the presumption against preemption. Indeed, his presumption is so strong that it can never be overcome absent express congressional intent or impossibility of compliance with both state and federal law. Thus, after *Wyeth*, a five-justice majority appears inclined to apply the presumption against preemption in obstacle preemption cases.

55. *See* Altria Group, Inc. v. Good, 129 S. Ct. 538, 555–56 (Thomas, J., dissenting) ("Since *Cipollone*, the Court's reliance on the presumption against pre-emption has waned in the express pre-emption context."). *Compare* Rowe v. N.H. Motor Transp. Assn., 552 U.S. 364 (2008), *Riegel*, 552 U.S. 312, Watters v. Wachovia Bank, 550 U.S. 1 (2007), Engine Mfrs. Assn. v. S. Coast Air Quality Mgmt. Dist., 541 U.S. 246 (2004), Sprietsma v. Mercury Marine, 537 U.S. 51 (2002), Geier v. Am. Honda Motor Co., 529 U.S. 861 (2000), United States v. Locke, 529 U.S. 89 (2000), *and* Freightliner Corp. v. Myrick, 514 U.S. 280 (1995) (failing to apply the presumption against preemption in the presence of an express preemption clause), *with Wyeth*, 129 S. Ct. at 1194–95, Pharm. Research & Mfrs. of Am. v. Walsh, 538 U.S. 644, 666 (2003), Rush Prudential HMO, Inc. v. Moran, 536 U.S. 355, 365–66 (2002), Johnson v. Fankell, 520 U.S. 911, 918 (1997), California v. ARC Am. Corp., 490 U.S. 93, 101 the presumption was absent from *Riegel v. Medtronic*,[56] where an express preemption clause revealed an unmistakable intent to preempt,[57] but was invoked with force in *Wyeth*, where congressional intent was far less certain. Of course, as the dissent in *Wyeth* points out, the presumption is not invoked in every implied preemption case. For instance, the presumption was notably absent in *Geier v. American Honda Motor Co.*[58] Even there, however, the presumption's use appears to be tied to congressional intent. *Geier* was ultimately decided on a theory of implied preemption,[59] but the statute at issue did contain an express preemption clause.[60] Congress had explicitly expressed its intent to preempt *something*; the Court was simply faced with the question of whether it might implicitly have intended to preempt other aspects of state law as well. Thus, while there was a danger of over-preempting within an area already the target of some preemption, there was no danger of preempting an area of law where Congress had intended no preemption whatsoever. In the absence of an express congressional intent to preempt, the Court is hesitant to infringe on areas of traditional state sovereignty, and it demonstrates that concern by its selective use of the presumption against preemption.

Finally, the Court's inconsistent standard of deference to agency determinations of preemption, like its inconsistent application of the presumption against preemption, appears to hinge on its concern for congressional intent and state sover-

---

(1989), *and* Hillsborough County v. Automated Med. Labs., Inc., 471 U.S. 707, 716 (1985) (applying the presumption in the context of implied preemption). *But see Altria Group*, 129 S. Ct. at 543; Bates v. Dow Agrosciences LLC, 544 U.S. 431, 449 (2005); *Lohr*, 518 U.S. at 485; Cipollone v. Liggett Group, 505 U.S. 504, 518 (1992) (applying the presumption in the context of express preemption).

56. 552 U.S. 312 (2008).

57. *Id.* at 316.

58. *Wyeth*, 129 S. Ct. at 1228–29 (Alito, J., dissenting) ("[T]he *Geier* Court specifically rejected the argument (again made by the dissenters in that case) that the 'presumption against pre-emption' is relevant to the conflict pre-emption analysis. Rather than invoking such a 'presumption,' the Court emphasized that it was applying 'ordinary,' 'longstanding,' and 'experience-proved principles of conflict pre-emption.'").

59. *Geier*, 529 U.S. at 866.

60. *Id.* at 867.

eignty.[61] The Court reserves its most deferential language for cases where congressional intent to preempt is clear. Where, on the other hand, congressional intent is less certain, the Court either neglects to mention agency views or treats them as useful only to the extent persuasive. Thus in *Medtronic v. Lohr*[62] the Court was "substantially informed by" the agency's view,[63] and in *Geier* the agency's position was entitled to "some weight,"[64] but in *Wyeth*, where an express preemption clause was lacking, the Court treated the agency's view as merely one among many potentially persuasive authorities.[65] When certain of congressional intent to preempt, the Court appears willing to accord substantial weight to agency views, even on questions that implicate federalism.

Two key observations emerge from this analysis. First, when considering issues of federal preemption, the Court is guided in large part by its respect for congressional intent and concern for the values of federalism. Second, the applicability of the presumption against preemption and the level of deference accorded to agency views is partly determined by the presence or absence of an express preemption clause. If the Court is convinced that Congress intended to preempt *something*, it perceives less need to take an active role in protecting state sovereignty, even if the precise scope of intended preemption is unclear. Thus, the deference agency views deserve depends on a sliding scale. Where preemptive intent is clear, agencies have some leeway to determine its 1188        *Harvard Journal of Law & Public Policy*        [Vol. 33scope, but where preemptive intent is unclear, agencies may not unilaterally make the preemptive determination.

---

61. Deference to agency determinations and the presumption against preemption are two sides of the same interpretive coin, and so it is unsurprising that both doctrines' applicability in a given case depends on the same considerations. If the presumption against preemption is accorded its full weight as a traditional tool of statutory construction capable of resolving textual ambiguities, statutes would rarely if ever be found to contain the ambiguously preemptive language necessary for *Chevron* deference to apply. The Court can either give *Chevron* its full weight or give the presumption its full weight, but not both. *See* Mendelson, *supra* note 2, at 745–46. Thus far the Court has adopted neither approach absolutely. Instead, when explicitly granting any weight to agency views, the Court has worked within the *Skidmore* framework, deferring selectively based on the agency's "thoroughness, consistency, and persuasiveness." *Wyeth*, 129 S. Ct. at 1201 (citing Skidmore v. Swift & Co., 323 U.S. 134, 140 (1944)).

62. 518 U.S. 470 (1996).

63. *Id.* at 495.

64. *Geier*, 529 U.S. at 883.

65. *Wyeth*, 129 S. Ct. at 1201.

This approach is both pragmatically desirable and theoretically justified. As many commentators have noted, a doctrine of across-the-board *Chevron* deference is ill-suited to preemption questions for a variety of reasons.[66] Agencies are specialized institutions with competence in core subject areas delegated by Congress, but they are unsuited to make general policy determinations outside those areas of expertise.[67] These concerns are heightened in the context of obstacle preemption, where congressional intent is least certain. If agencies are inexpert at determining preemptive *scope* in the presence of an express preemption clause, how much more inexpert must they be at determining, in the absence of an express preemption clause, whether there is to be any preemption at all? Furthermore, as Justice Thomas emphasized in his concurrence, the Constitution empowers Congress,[68] not executive agencies or the courts, to preempt state law.[69] If state law is to be displaced, that decision must be based on something more than a court or agency determination that state law poses an obstacle to vague congressional objectives.

---

66. *See, e.g.*, Mendelson, *supra* note 2, at 779–97 (citing concerns of agency inexpertness, arbitrary decision making, and self-aggrandizement); Catherine M. Sharkey, *What* Riegel *Portends for FDA Preemption of State Law Products Liability Claims*, 103 NW. U. L. REV. 437, 448 (2009) ("Sweeping *Chevron* deference to agencies on preemption questions raises the troubling specter of enabling or encouraging cycles of agency political flip-flop, and, more generally, of forgoing a key judicial check by relieving the agency of responsibility to supply an adequate record to substantiate its position regarding the preemptive effect of federal statutes and regulations.").

67. *See* Mendelson, *supra* note 2, at 780–82.

68. U.S. CONST. art. VI, cl. 2 ("This Constitution, and the Laws of the United States . . . shall be the supreme Law of the Land; and the Judges in every State shall be bound thereby, any Thing in the Constitution or Laws of any State to the Contrary notwithstanding.").

69. *Wyeth*, 129 S. Ct. at 1207 (Thomas, J., concurring in the judgment) ("[A]gency musings . . . do not satisfy the Art. I, § 7 requirements for enactment of federal law and, therefore, do not pre-empt state law under the Supremacy Clause. When analyzing the pre-emptive effect of federal statutes or regulations validly promulgated thereunder, '[e]vidence of pre-emptive purpose [must be] sought in the text and structure of the [provision] at issue' to comply with the Constitution." (quoting CSX Transp., Inc. v. Easterwood, 507 U.S. 658, 664 (1993))).

Electronic copy available at: https://ssrn.com/abstract=1621325

Thus, the *Chevron-Mead* framework,[70] or even an across-the-board doctrine of lower, *Skidmore*-style deference to agency determinations, is inappropriate.[71] Such frameworks would accord the same level of deference to agency views whether or not Congress authorized such preemption in the form of an express preemption clause. Rather, deference should be tailored to suit the individual case. Where Congress, by use of an express preemption clause, clearly intended to preempt something, and the question is merely one of preemptive scope, some level of deference—*Skidmore*, or perhaps even *Chevron*—is appropriate.[72] Where, however, intent to preempt is not obvious, courts should reserve the question for Congress by invoking the presumption against preemption.

*Gregory M. Dickinson*

---

70. *See The Supreme Court, 2008 Term—Leading Cases*, 123 HARV. L. REV. 153, 271 (2009) (arguing that the *Chevron-Mead* framework can adequately account for values of federalism). Even when the additional nuances of *United States v. Mead Corp.*, 533 U.S. 218 (2001), are added, the framework remains incapable of varying deference based on congressional intent to preempt. *Mead*'s focus is on formality of agency procedure, at best a poor proxy for preemptive intent. An agency is quite often empowered to utilize formal procedures even where Congress intends its enactments to have no preemptive effect.

71. A number of commentators advocate across-the-board *Skidmore* deference in preemption cases. *See, e.g.*, Mendelson, *supra* note 2, at 797–98; Sharkey, *supra* note 2, at 491–98.

72. The argument that deference to agency preemption determinations should vary based on the presence or absence of an express preemption clause is also supported by the rationale underlying *Chevron* deference. The *Chevron* doctrine asks courts to determine whether Congress has delegated interpretive authority to a given agency. If a statute is ambiguous on a particular point, Congress is presumed to have delegated authority to make that determination to the agency responsible for administering the statute. Thus, where Congress has imprecisely spoken via an express preemption clause, *Chevron* counsels deference to an agency's reasonable interpretation. Where, however, Congress has not spoken at all—where the question is one of implied preemption—Congress should not be presumed to have delegated such important decision-making authority to an agency.